\documentclass[aps, prl, twocolumn, groupedaddress, showpacs, floatfix]{revtex4}
\usepackage{amsmath, amssymb}
\usepackage[english]{babel}
\usepackage[latin1]{inputenc}
\usepackage{xy}
\usepackage{graphicx}
\usepackage{verbatim}
\usepackage{enumerate}
\usepackage{url}
\usepackage{subfigure}
\usepackage{float}
\usepackage{dcolumn}% Align table columns on decimal point
\usepackage{bm}% bold math
\usepackage{times}

\newcommand{\Z}{\mathbb{Z}}

\newcommand{\R}{\mathbb{R}}
\newcommand{\N}{\mathbb{N}}

\newcommand{\Cc}{\mathcal{C}}

\DeclareMathOperator{\Stab}{Stab}

\newcommand{\Sym}[1]{S_{#1}}
\newcommand{\ud}{\mathrm{d}}

\newcommand{\rset}[2]{\left\lbrace #1\ \left|\ #2\right.\right\rbrace}

\newcommand{\set}[2]{\rset{#1}{#2}}

\newcommand{\abs}[1]{\left| #1 \right|}

\newcommand{\ind}[1]{_{\mathrm{#1}}}
\newcommand{\vphi}{\varphi}

\begin{document}

\preprint{APS/123-QED}

%%%%%%%%%%%%% TITLE

\title{Chaos in Symmetric Phase Oscillator Networks}% Force line breaks with \\
\author{Christian Bick${}^{a,b}$}%
\author{Marc Timme${}^{a,c}$}
\author{Danilo Paulikat${}^c$}
\author{Dirk Rathlev${}^c$}
\author{Peter Ashwin${}^d$}%

\affiliation{%
${}^a$Network Dynamics Group, Max Planck Institute for Dynamics and Self-Organization (MPIDS), 37073 G\"ottingen, Germany\\
${}^b$Institute for Mathematics, Georg--August--Universit\"at G\"ottingen, 37073 G\"ottingen, Germany\\
${}^c$Faculty of Physics, Georg--August--Universit\"at G\"ottingen, 37077 G\"ottingen, Germany\\
${}^d$Mathematics Research Institute, University of Exeter, Exeter EX4 4QF, United Kingdom}%

\date{\today}

%%%%%%%%%%%% ABSTRACT

\begin{abstract}
Phase-coupled oscillators serve as paradigmatic models of networks of weakly
interacting
oscillatory units in physics and biology. The order parameter which quantifies
synchronization was so far found to be chaotic only in systems with inhomogeneities.
Here we show that even symmetric systems of identical oscillators may not only
exhibit chaotic dynamics, but also chaotically fluctuating order parameters. 
Our findings imply that neither inhomogeneities nor amplitude variations are
necessary to obtain chaos, i.e., nonlinear interactions of phases give rise
to the necessary instabilities.
\end{abstract}

\maketitle

%%%%%%%%%%%% INTRODUCTION

%\section{Introduction}
\textit{Introduction\ ---\ }
Models of coupled oscillators describe various collective phenomena in
natural and artificial systems, including the synchronization of flashing
fireflies, or superconducting Josephson junctions, oscillatory neural activity
and oscillations in chemical reaction kinetics \cite{Pikovsky2003}. %, Strogatz2004}.
In particular, phase-coupled models serve as paradigmatic approximations
for many weakly coupled limit cycle oscillators \cite{Ashwin1992, Swift1992}.
The Kuramoto model (and its extensions) provides the gold standard in this
field because it suitably describes the dynamics of a variety of real
systems, is extensively studied numerically and reasonably understood
analytically \cite{Kuramoto, Acebron2005}. Each oscillator $k$ with
phase $\vphi_k(t) \in \R/2\pi\Z =: T$ on the 1-torus $T$ changes with time
$t$ according to
\begin{equation}\label{eq:Model}
\frac{\ud\vphi_k}{\ud t}=\omega_k+\frac{1}{N}\sum_{j=1}^{N}g(\vphi_k-\vphi_j)
\end{equation}
for all $k\in\{1, \ldots, N \}$. For the original Kuramoto model the coupling
function $g$ has a single Fourier mode, $g=\sin$.
The dimension of such systems can be reduced to low dimensions
\cite{Watanabe1993, Ott2008, Marvel2009}, implying dynamics that
is either periodic or quasi-periodic.
For coupling functions with two or more Fourier components the collective
dynamics may be much more complicated.
For example, stable heteroclinic switching may emerge \cite{Ashwin2007,
Ashwin2010a}. More irregular, chaotic dynamics of system \eqref{eq:Model}
is observed for non-identical oscillators only
\cite{Popovych2005, Luccioli2010}, raising the question whether inhomogeneities
are necessary for the occurrence of such dynamics. To the best of our
knowledge, the only hint that chaotic dynamics might exist for symmetric
phase-oscillator networks are attractors with irregular structure in phase space
found recently in a system of $N=5$ oscillators \cite{Ashwin2007}.

The complex order parameter
\begin{equation}\label{eq:OrderParam}
R(t) = \frac{1}{N}\sum_{j=1}^{N}\exp(i\varphi_j(t))\ \in \mathbb{C}
\end{equation} where $i=\sqrt{-1}$
constitutes an important characteristic for coupled oscillator systems. 
In particular, its absolute value $\abs{R(t)}$ quantifies their
degree of synchrony with $\abs{R(t)}=1$ if all oscillators are
in phase. For the original Kuramoto system the full complex order
parameter \eqref{eq:OrderParam} acts as a mean field variable enabling
closed-form analysis \cite{Strogatz2000}.

In homogeneous, globally coupled systems \eqref{eq:Model} it remains
unknown whether or not there exists any coupling function $g$ that
gives rise to chaotic order parameter fluctuations. Synchronous solutions,
anti-synchronous splay states as well as the dynamics of cluster states
(where $\vphi_k=\vphi_j$ for at least one $k\neq j$) all yield a periodic
complex order parameter $R(t)$. For any invariant solutions on tori,
$R(t)$ is either periodic or quasiperiodic. The most irregular dynamics
of $R(t)$ observed so far is due to heteroclinic cycles where $R(t)$
is non-periodic as it `slows down' each cycle. Even if chaotic dynamics
does emerge within the system, it may average out due to symmetry,
possibly resulting in a regular dynamics of the order parameter.

In this Letter, we answer the question whether chaotic dynamics, and
moreover, chaotic order parameter fluctuations may arise for some $g$
even in the absence of inhomogeneities in~\eqref{eq:Model}. 
We show that indeed chaos is not
possible for $N<4$. By contrast, for $N = 4$, chaotic attractors
can appear in a specific family of coupling functions $g$. Interestingly,
attractors of all theoretically possible symmetries exist and we provide
further examples of attracting chaos for $N = 5$ and $N = 7$. The existence
of chaos for infinite families of $N\geq 4$ implies that chaos occurs in
systems with certain $N>N_0$ for any $N_0\in\N$ and suggests that chaos is
likely to occur in many high-dimensional systems with a suitable choice of~$g$.

%%%%%%%%%%%% MODEL

\smallskip\textit{No chaos for $N=2$ and $N=3$ \ ---\ }
Let $T^N$ be the $N$-dimensional torus and let $\Sym{N}$
denote the group of permutations of $N$ symbols. Suppose $M$ is a
differentiable manifold and let $\Gamma$ be a group that acts on $M$.
Recall that a vector field $X$ on $M$ is called $\Gamma$-equivariant
if $X$ `commutes' with the action of $\Gamma$, i.e., $X\circ \gamma
= \hat{\gamma}\circ X$ for all $\gamma\in\Gamma$ where $\hat{\gamma}$ denotes
the induced action on the tangent space.

Equivariance implies restrictions of the dynamics specified by the
vector field. We study the dynamical system on $T^N$ given by the
ordinary differential equations \eqref{eq:Model}. Let us henceforth
assume that the system is homogeneous, i.e., $\omega_k=\omega$ for
all $k \in \{1, \ldots, N\}$. This system is $\Sym{N}\times T^1$ equivariant
where $\Sym{N}$ acts by permuting indices and $T^1$ through a
phase shift. Recall some basic properties of this system
\cite{Ashwin1992}. Introducing phase differences 
$\psi_{j}:=\vphi_{j}-\vphi_1$ for all $j\in\{1, \ldots, N\}$ eliminates the
phase-shift symmetry. Write $\vphi=(\vphi_1,
\ldots, \vphi_N)$ and $\psi=(\psi_1, \ldots, \psi_N)$. The reduced
system on $T^{N-1}$ is given by 
\begin{equation}
\dot\psi_j=\frac{1}{N}\left(\sum_{k=1}^{N}g(\psi_j-\psi_k)-\sum_{k=1}^{N}g(-\psi_k)\right)
\label{eq:PsiDynamics}
\end{equation}
for all $j\in\{2, \ldots, N\}$.

For any partition $P = \{P_1, \ldots, P_m\}$ of $\{1, \ldots, N\}$ (that
is $P_r\subset \{1, \ldots, N\}$, $\bigcup_{j=1}^m P_j =  \{1, \ldots, N\}$,
and $P_r\cap P_s=\emptyset$ for $r\neq s$) the subspaces
%\[
\begin{equation}
%F_P := \set{\vphi}{ \forall\ P_r\in P:\ \vphi_j=\vphi_k\ \forall\ (j, k)\in P_r^2  }\subset T^{N}
%F_P := \set{\vphi}{\vphi_j=\vphi_k\ \text{whenever there is }r\text{ s.th. }j, k\in P_r}\subset T^{N}
F_P := \set{\vphi}{j, k\in P_r \text{ for any }r \implies \vphi_j=\vphi_k}\subset T^{N}
%\label{eq:Subspaces}
\end{equation}
%\]
are flow-invariant. The subspaces divide $T^{N-1}$ in $(N-1)!$ invariant
$(N-1)$-dimensional simplices \cite{Ashwin1992}; one of which
%\[
\begin{equation}
\Cc := \set{\psi}{0=\psi_1<\psi_2 < \ldots < \psi_{N} < 2\pi}\subset T^{N-1}
%\label{eq:Simplex}
\end{equation}
%\]
we refer to as the \emph{canonical invariant region}. There is
a $\Z_N := \Z/N\Z$ symmetry on the canonical invariant region and
the `splay state' (the phase-locked state with $\psi_j = 2\pi j/{N}$)
is the only fixed point of this action at the centroid of this region.

The reduction of symmetry has implications for the existence of
chaos in low dimensions. For $N=2$ and $N=3$ the phase space of the reduced
system is a one, resp.~two-dimensional torus. This means that by the 
Poincar\'e--Bendixon theorem \cite{Schwartz1963} chaos is not possible
in these systems for $N<4$.

%%%%%%%%%%%% CHAOS FOR N=4

\begin{figure}%[h]%
\includegraphics[scale=0.77]{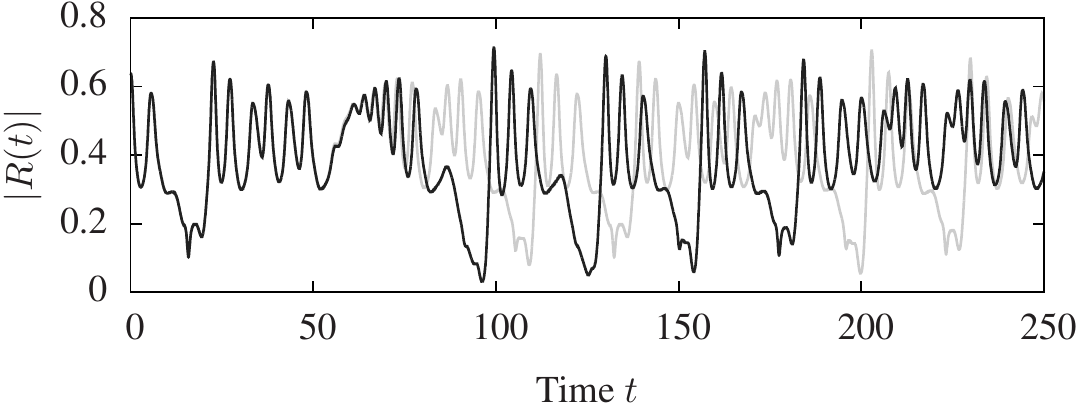}
\caption{\label{fig:MeanField}Trajectories of the absolute value of the
order parameter $\abs{R(t)}$ fluctuate
chaotically ($N=4$). Two trajectories with small difference in initial condition
diverge from each other  (coupling function \eqref{eq:2dg} for parameter values
$\eta_1=0.1104$ and $\eta_2=0.5586$).}
\end{figure}

\begin{figure}[t]
\includegraphics[scale=.79]{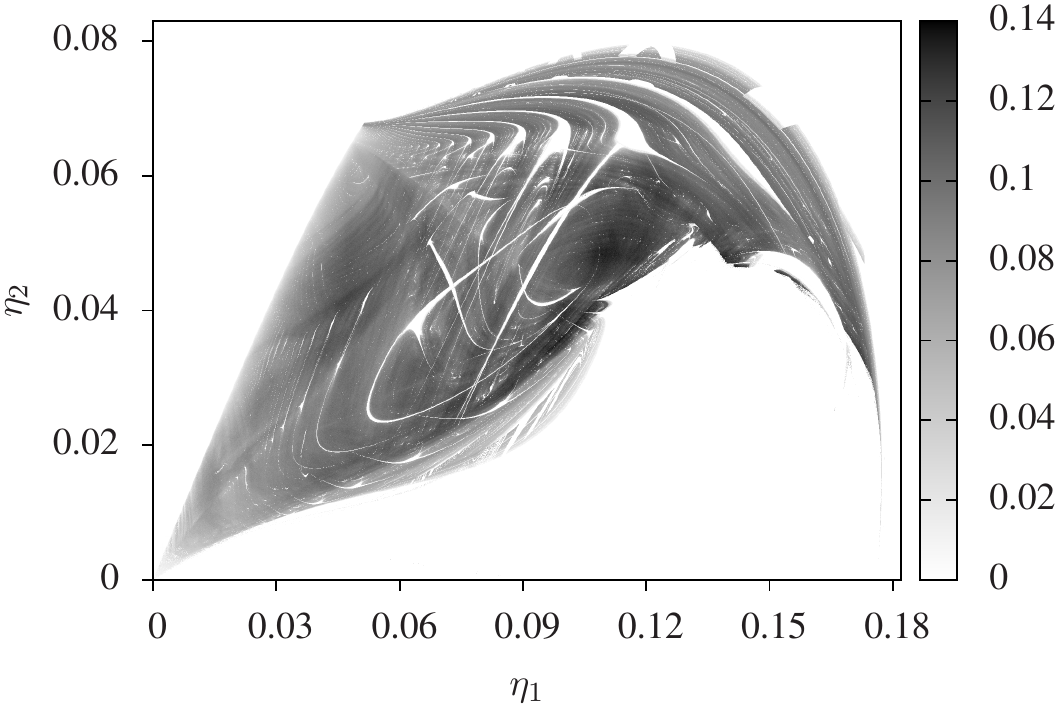}
\caption{\label{fig:Sweeps} Chaos for $N=4$. Maximal Lypunov exponent
is positive in a region of parameter space. The initial condition
was fixed and the coupling function parametrized by \eqref{eq:2dg}.}
%to $\vphi(0)=(6.1821, 1.3693, 1.9405, 2.5897)$
\end{figure}

\smallskip\textit{Chaos and symmetry for $N = 4$\ ---\ }
We choose a parametrization of the coupling function $g$ in \eqref{eq:Model}
by considering a truncated Fourier series
%\[
\begin{equation}
g(\vphi) = \sum\limits_{k=1}^4a_k\cos(k\vphi+\xi_k).
%\label{eq:fourFourier}
\end{equation}
%\]
In particular, we restrict ourselves to the two parameter family given
by the parametrization
\begin{equation}\label{eq:2dg}
(\xi_1, \xi_2, \xi_3, \xi_4) = (\eta_1, -\eta_1, \eta_1+\eta_2, \eta_1+\eta_2)
\end{equation}
where $\eta_1$ and $\eta_2$ are real valued parameters and $a_1=-2, a_2=-2,
a_3=-1,$ and $a_4=-0.88$ are constants.

\begin{figure*}
\subfigure[$\ \Sigma(A) = 1$]{\includegraphics[scale=.73]{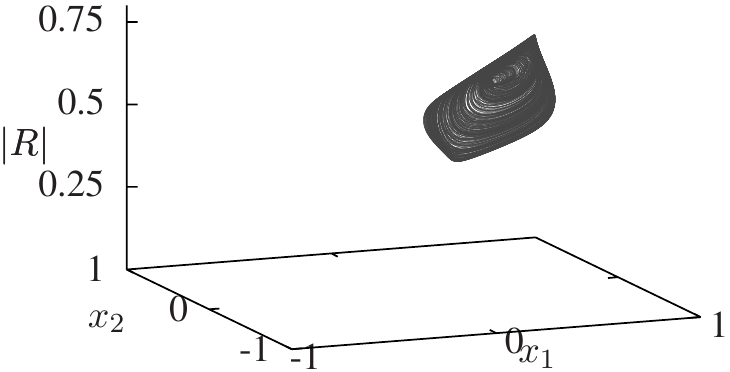}}\quad\quad
\subfigure[$\ \Sigma(A) = \Z_2$]{\includegraphics[scale=.73]{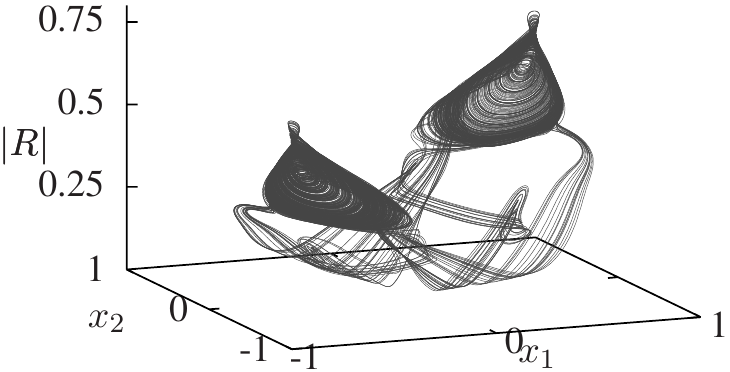}}\quad\quad
\subfigure[$\ \Sigma(A) = \Z_4$]{\includegraphics[scale=.73]{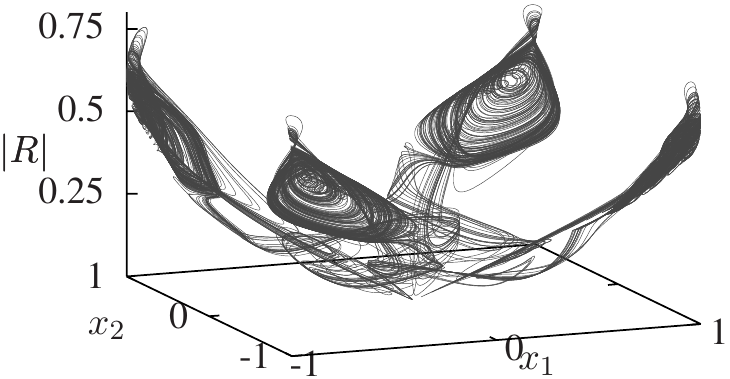}}
\caption{\label{fig:AttractorSymmetry}All possible symmetries of the
chaotic attractors for different parameters for $N=4$. We have $\eta_1=0.138$ in panel (a), $\eta_1=0.0598$ in panel
(b), $\eta_1=0.1104$ in panel (c), and $\eta_2=0.5586$ in all panels. 
The projection is a $\Gamma=\Sym{4}$ equivariant map, $x_1 = \sin(\vphi_2-\vphi_4),
x_2 = \sin(\vphi_1-\vphi_3),$ and $x_3=\abs{R}$.}
\end{figure*}

For $N=4$, chaotic attractors do indeed exist. The dynamics of the absolute
value of the order parameter exhibits chaotic fluctuations and exponential
divergence of trajectories, cf.~Figure~\ref{fig:MeanField}. To explore
parameter space, we calculated the maximal Lyapunov exponent
$\lambda\ind{max}$ from the variational equations \footnote{The time step
of the standard Runge--Kutta scheme was typically chosen to be
$\Delta t=0.05$ and integration ranged over several thousand time
units.}. There are regions in $(\eta_1, \eta_2)$-parameter space in
which $\lambda\ind{max}$ is greater than zero, cf.
Figure~\ref{fig:Sweeps}. As might be expected, there is fine structure in this region,
for example islands where the trajectory converges to a stable limit
cycle. Lines of period doubling cascades \footnote{Numerical continuation
of the bifurcations was done with AUTO-07p, E.~J.~Doedel \emph{et.~al.} (indy.cs.concordia.ca/auto/).} bound the chaotic region and
end in a homoclinic flip bifurcation with an inclination flip
\cite{Homburg2000} (details not shown).
Exploring initial conditions revealed the coexistence of chaotic
attractors and stable limit cycles in part of the chaotic region.

Chaotic attractors in equivariant dynamical system can exhibit
symmetries themselves. Let $A$ be a chaotic attractor as defined in
\cite{Golubitsky2002}, i.e., a Lyapunov-stable, closed, and connected
set that is the $\omega$-limit set of a trajectory, for a dynamical
system on a manifold $M$ given by a $\Gamma$-equivariant vector field.
The subgroup $\Stab(A) := \set{\gamma\in \Gamma}{\gamma(a)=a\ \text{for all}\  a\in A}$ is the
group of instantaneous symmetries of the attractor, i.e., at any point
in time the action of $\Stab(A)$ keeps every point in $A$ fixed.
Furthermore, we define $\Sigma(A) := \set{\gamma\in\Gamma}{\gamma(A)=A}$ to be the set of
symmetries on average, and we have $\Stab(A)\subset\Sigma(A)$ as a
subgroup.

The subdivision of the phase space by flow-invariant regions restricts the
possible symmetries of chaotic attractors. The possible symmetries on average
of any chaotic attractor $A\subset \Cc$ with trivial instantaneous
symmetries $\Stab(A)=\{1\}$ are limited to subgroups of $\Z_N$ since
they are contained one of the invariant simplices ($\Cc$ or one of its
images under the group action) with that symmetry. For $N=4$, any chaotic
attractor of this type must have trivial instantaneous symmetry. Thus,
the possible symmetries on average are limited to subgroups of $\Z_4$, i.e., 
$\Sigma(A)\subset\Z_4$. In fact, we have found examples of chaotic
attractors for each possible symmetry in systems of $N=4$ and coupling
functions given by \eqref{eq:2dg} (Figure~\ref{fig:AttractorSymmetry}).
Note that this definition of attractor is somewhat restrictive---Milnor
attractors may display a wider range of symmetries including different
instantaneous symmetries at the same time.

%%%%%%%%%%%% CHAOS FOR N>4

\begin{figure}[b]
\includegraphics[scale=.8]{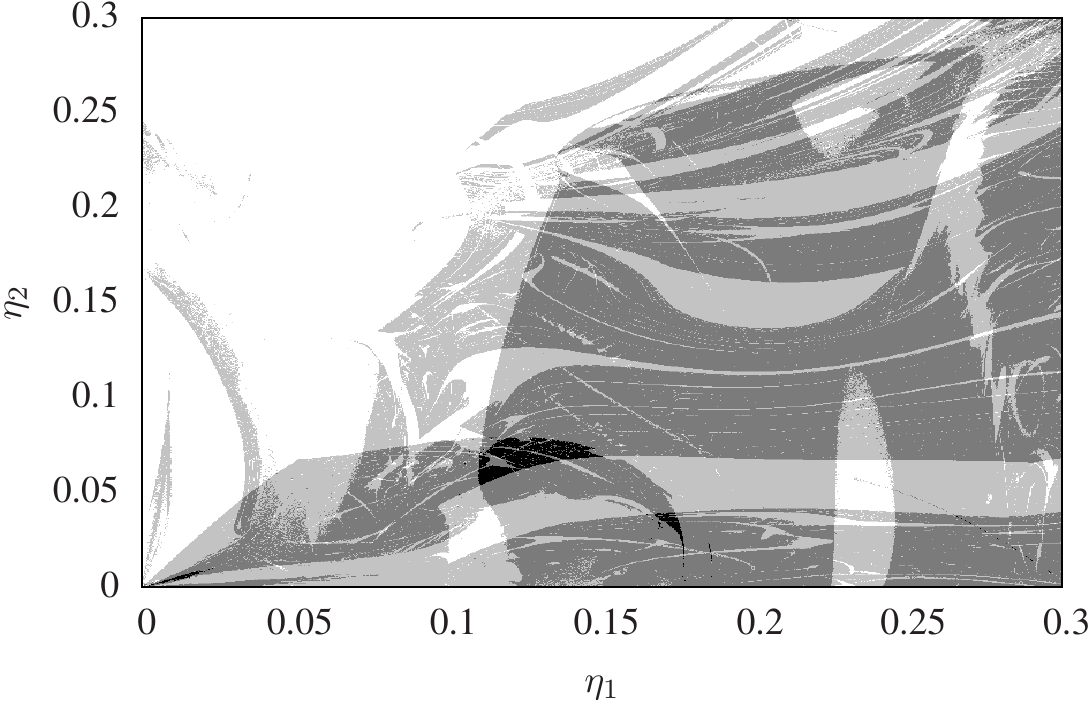}
\caption{\label{fig:MultipleN}Overlapping regions where the maximal
Lyapunov exponent is greater than $0.01$ for $N \in \{4,5,7\}$. The darker the
color, the more $N$ for which the condition holds. For parameter values
around
$(0.115, 0.06)$ there is a region where there is chaos for all
these three $N$.}
\end{figure}

\smallskip\textit{Chaos for $N > 4$\ ---\ }
Analyzing the same region of parameter space for $N>4$ yields attracting
chaos in systems of $N = 5$ and $N=7$ oscillators in large regions.
Figure~\ref{fig:MultipleN} shows an overlay of regions 
for three different $N$; regions are shaded where the Lyapunov exponent
exceeds $0.01$ and darker areas indicate that several $N$ satisfy
this condition. Clearly, there is a single coupling function for which
attracting chaos is present for all $N=4$, $N=5$ and, $N=7$. 
Intriguingly, we did not find chaotic attractors for any $N\in \{6, 8,
9,\ldots, 13\}$ in the entire region of parameter space considered
in Fig.~\ref{fig:MultipleN}.

The parametrization of the coupling function by a truncated Fourier
series raises the question how many Fourier components the coupling
function needs to contain for chaos to occur. For $N=5$ we also measured
positive Lyapunov exponents when the coupling was chosen to be through
the simpler coupling function $g(\vphi)=-0.2\cos(\vphi + \eta_1) - 0.04\cos(2\vphi - \eta_2)$
as in \cite{Ashwin2007}. Hence, in dimension five, coupling functions
with only two Fourier components suffice to generate chaotic dynamics
whereas for $N=4$, we did not find an example with less than four components.

\begin{figure}[b]
\includegraphics[scale=.8]{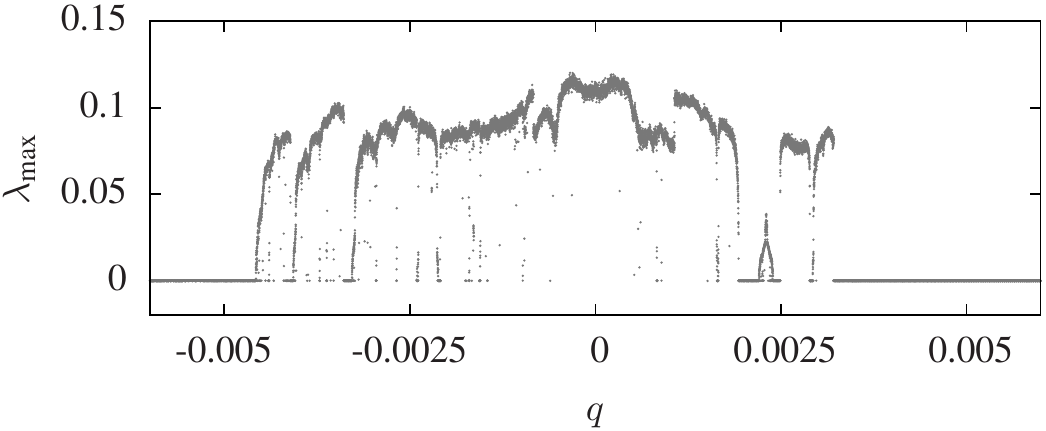}
\caption{\label{fig:ChaoticSaddles}Positive maximal Lyapunov exponents
for asymmetric $4$-cluster states for large systems, $N=KM+L(q, M)\gg 4$.
Here $q$ parametrizes the deviation from the symmetric cluster state and
$L$ is the corresponding integer dimension (coupling function \eqref{eq:2dg}
with $\eta_1=0.1104$ and $\eta_2=0.5586$). }
\end{figure}

From the above, it is clear that for systems of size $N=KM$ with 
$K\in \{4, 5, 7\}$ there are chaotic invariant sets lying in flow-invariant
subspaces for coupling functions yielding positive $\lambda\ind{max}$,
cf.~Figure~\ref{fig:MultipleN}. For instance, for $K=4$, these spaces
are given by partitions $P=\{P_1, \ldots, P_4\}$ with $\abs{P_j}=M$
for $j \in \{1, \ldots, 4\}$. For $N$ large we similarly calculated a
positive maximal Lyapunov exponents for the system reduced to asymmetric
$4$-cluster states given by partitions $P=\{P_1, \ldots, P_4\}$ with
$\abs{P_1}/N=1/4+q$ and $\abs{P_j}/N=1/4-q/3$ for $j\in \{2, 3, 4\}$
as depicted in Figure~\ref{fig:ChaoticSaddles}. However, these chaotic
invariant sets in invariant subspaces close to the symmetric cluster
state may be transversally repelling, possibly yielding non-chaotic
long-term dynamics.

%%%%%%%%%%%% DISCUSSION

%\section{Discussion}
\smallskip\textit{Discussion\ ---\ }
Inhomogeneities or asymmetries are thus not necessary for collective chaotic dynamics to
appear in system \eqref{eq:Model}, and even chaotic order parameter
fluctuations emerge in the presence of full $S_N\times T$-symmetry. We
highlighted that for certain coupling functions
chaotic attractors exist for several $N$. However, the regions in
parameter space for which chaotic attractors exist vary drastically,
cf.~Figure~\ref{fig:MultipleN}. For certain coupling functions there
are chaotic invariant sets lying in flow-invariant
subspaces that correspond to the symmetric and near-symmetric cluster
states but these may not be transversally attracting. The question
remains whether there are coupling functions giving rise to
chaotic sets that are actually attracting for $N=6$ and $N\geq 8$.
Moreover, is there a `universal chaos function' in the sense that
there is a coupling function for which there is some $N_0\in\N$
such that there exists a chaotic attractors for all (or at least an
infinite number of) $N>N_0$?

For coupling functions with only one Fourier component, finite
dimensional systems and the continuum limit are related; the
dynamics for both finite $N$ and in the continuum limit reduces to
effectively two-dimensional dynamics \cite{Ott2008, Marvel2009}
preventing the occurrence of
chaotic trajectories. Is chaos possible in the continuum limit for
more complicated coupling functions? If so, how would such a result
relate to chaos in the finite-dimensional systems we
have studied here?

For finite systems, attracting chaos in the system does not necessarily
imply chaotic dynamics of the order parameter since there could be
chaotic fluctuations for example in an `antiphase state.' The converse,
however, holds. Additionally, observed chaotic fluctuations of the
order parameter cannot necessarily be traced back to inhomogeneity
in the system (possibly through an experimental setup, cf.~\cite{Kori2008})
because, as shown above, even fully symmetric systems can support such
dynamics. When considering the continuum limit, the problem of these
implications becomes more subtle and will require further investigation.

The coupling function we considered above are written in terms of a
truncated Fourier series. As discussed above, the number of Fourier
components is relevant for the dynamics. An alternative approach
would be to consider suitable piecewise linear functions. For $N=4$,
we find that systems with piecewise linear $g$ also exhibit positive
maximal Lyapunov exponents (not shown). Finding a suitable basis
for the space of coupling functions might be a way to explain some
of the dynamical features that were observed.

Coupled phase oscillators are a limit of weakly coupled limit cycle
oscillators \cite{Ashwin1992}. In globally coupled identical 
Ginzburg--Landau oscillator ensembles, chaotic dynamics can be
observed \cite{Hakim1992, Nakagawa1994}. However, it was thought that
the amplitude degree of freedom is crucial for the emergence of such
dynamics. Our results
show that this is not the case and, moreover, suggest that
chaotic mean field oscillations are also present in a large class
of higher-dimensional symmetrically coupled limit cycle oscillators
with a rich possible range of chaotic dynamics.
%However, a general
%way to extrapolate from the coupling function in the limiting case
%to find chaos in a generic setting is yet to be discovered.

%%%%%%%%%%%% ACKNOWLEDGEMENTS
%\section{Acknowledgements}
\smallskip\textit{Acknowledgements\ ---\ }
CB would like to thank Laurent Bartholdi. Supported by the Federal
Ministry for Education and Research (BMBF) Germany under grant number
01GQ1005B, a grant by the NVIDIA Corporation, USA, and a grant of the
Max Planck Society to MT.

%%%%%%%%%%%% BIBLIOGRAPHY

\bibliographystyle{apsrev}
\def\urlprefix{}
\def\url#1{}

%\bibliography{/Volumes/Media/Chris/Uni/Papers/BibTeX/library} %home
%\bibliography{/home/chris/Papers/BibTeX/library} %ucsd
%\bibliography{/home/bick/Papers/BibTeX/library} %mpi
\bibliography{prl_oscchaos} %mpi

\end{document}